\documentclass[journal]{IEEEtran}
\usepackage{cite}
\usepackage{amsmath,amssymb,amsfonts}
\usepackage{algorithmic}
\usepackage{graphicx}
\usepackage{textcomp}

\ifCLASSINFOpdf
\else
\fi

\begin{document}
%
\title{Quasioptic, Calibrated, Full 2-port Measurements of Cryogenic Devices under Vacuum in the 220 - 330 GHz Band}

%
%
%

\author{Maxim~Masyukov,
        Aleksi~Tamminen,
        Irina~Nefedova,
        Andrey~Generalov,
        Samu-Ville~Pälli,
        Roman Grigorev,
        Pouyan Rezapoor,
        Rui Silva,
        Juha Mallat,
        Juha Ala-Laurinaho
        and~Zachary~Taylor,~\IEEEmembership{Member,~IEEE}
\thanks{The work has been supported by the Finnish Foundation for Technology Promotion and the Magnus Ehrnrooth Foundation and has been performed at MilliLab, the European Space Agency's external laboratory.}

\thanks{M. Masyukov (corresponding author, email: maxim.masyukov@aalto.fi), A. Tamminen, I. Nefedova, S.-V. Pälli, R. Grigorev, P. Rezapoor, J. Mallat and Z. Taylor are with the MilliLab, Department of Electronics and Nanoengineering, Aalto University, P.O. Box 15500, 00076 Aalto, Finland. }

\thanks{Andrey Generalov is with VTT Technical Research Centre of Finland, P.O. Box 1000, FI-02044 VTT, Finland.}
\thanks{ R.  Silva is with Center for Microelectromechanical System (CMEMS-UMinho), University of Minho, Campus de Azurém, 4800-058 Guimarães, Portugal.}
\thanks{Manuscript received XX.XX.XXXX; revised XX.XX.XXXX.}}

%
%

\markboth{IEEE Transactions on Terahertz Science and Technology,~Vol.~N, No.~N, \today }%
{Masyukov \MakeLowercase{\textit{et al.}}:Quasioptic, Calibrated, Full 2-port Measurements of Cryogenic Devices under Vacuum in the 220 - 330 GHz Band}
%



\maketitle

\begin{abstract}
 
A quasi-optical (QO) test bench was designed, simulated, and calibrated for characterizing S-parameters of devices in the 220–330 GHz (WR-3.4) frequency range, from room temperature down to ~ 4.8 K. The devices were measured through vacuum windows via focused beam radiation. A de-embedding method employing line-reflect-match (LRM) calibration was established to account for the effects of optical components and vacuum windows. The setup provides all four S-parameters with the reference plane located inside the cryostat, and achieves a return loss of $\ approx$30 dB with an empty holder. System validation was performed with measurements of cryogenically cooled devices, such as bare silicon wafers and stainless-steel frequency-selective surface (FSS) bandpass filters, and superconducting bandpass FSS fabricated in niobium. A permittivity reduction of Si based on 4-GHz resonance shift was observed concomitant with a drop in temperature from 296 K to 4.8 K. The stainless steel FSS measurements revealed a relatively temperature invariant center frequency and return loss level of 263 GHz and 35 dB on average, respectively. Finally, a center frequency of 257 GHz was measured with the superconducting filters, with return loss improved by 7 dB on average at 4.8 K. To the best of our knowledge, this is the first reported attempt to scale LRM calibration to 330 GHz and use it to de-embed the impact of optics and cryostat from cryogenically cooled device S-parameters.

\end{abstract}


%
\IEEEpeerreviewmaketitle

\section{\label{sec:introduction}Introduction}

Recent advancements in enhancing millimeter- and sub-millimeter-wave / terahertz (THz) detection instruments for a variety of missions—whether conducted on the ground, via balloons, or in orbit—have resulted in significant collaborative achievements\cite{jovanovic20232023}. The frequency band of 100 GHz - 10 THz is a crucial domain hosting spectral lines that serve as indicators of star and planet formation, the evolution of matter in galaxies, and the intricate astrochemistry of interstellar clouds\cite{kulesa2011terahertz,kermish2012polarbear}.

Orbital missions like Odin\cite{odin} and SWAS\cite{franklin2008swas} have featured receivers utilizing cutting-edge room-temperature or cooled Schottky technology. Launched in 2009 by the European Space Agency, the Herschel Space Observatory\cite{pilbratt2010herschel} and the Planck satellite\cite{mandolesi2002planck} have played a substantial role in advancing our understanding of the universe at THz frequencies by employing cryogenically cooled superconducting instruments\cite{siegel2007thz}. Herschel, equipped with a sizable telescope, conducted detailed observations of cold, dark celestial regions, providing new insights into star formation and galaxy structure\cite{jackson2006low}. Meanwhile, Planck, focused on cosmic microwave background radiation, meticulously mapped the universe's oldest light, offering precise measurements that enriched our comprehension of the early cosmos and the fundamental parameters of the Big Bang model. These missions, collectively, have significantly elevated our insights into astrophysical processes and the cosmic microwave background across the THz spectrum\cite{bianchini2014cross}.

A key area of support for these instruments and their development is comprehensive, lab-based qualification of components including electromagnetic performance of quasioptical devices at low temperatures. Thus, complex experimental techniques for S-parameter characterization and noise measurements of cryogenically cooled devices, at THz frequencies, are frequently employed. However, conventional waveguide-based approaches for characterizing THz devices face limitations stemming from signal attenuation, bandwidth restrictions, and the challenging task of implementation in a cryogenic environment\cite{anferov2020millimeter}. Thus, while there are significant advances in, e.g., using novel superconducting technologies suitable for higher frequencies \cite{kim2024wafer}, there remains significant challenges for calibration and measuring techniques at low temperatures beyond ~ W-Band (110 GHz), especially when one tries to translate the microwave-based technology into millimeter waves\cite{yeh2013situ,wang2021cryogenic}.

Quasioptical approaches for coupling electromagnetic waves to cryogenically cooled devices are one way to reduce complexity and allow the measurement transceiver electronics to remain outside the cryostat. However, shifting the reference plane inside the cryostat to the DUT is challenging, and not all conventional calibration standards are compatible with vacuum and/or cryogenic environments. Further, repeated standards placement and eventual DUT mounting may require numerous vacuum breaks which can have significant detrimental effects on extracting the error network; especially if parts need to be disassembled. Thus, a calibration approach with minimal interference to the quasioptical or waveguide signal path is crucial. Quasioptical calibration for shifting the reference plane to the DUT surface has been studied at room temperatures\cite{legg2013implementation,bourreau2006quasi} with Thru, Reflect, and Line calibration standards, however, this approach cannot be used with vacuum windows since de-embedding requires one to change the quasioptical path between the vacuum windows by the required line length change. In place of TRL, a Line-Reflect-Match (LRM) calibration approach can be used, which does not set limitations on shifting the reference planes for line-type standards.

In this paper, a methodology is proposed for measuring all four S-parameters of passive, two-port, cryogenically cooled devices in the 220–330 GHz range. This approach keeps the VNA and optics at room temperature, requires only the DUT inside the cryostat, and enables de-embedding of DUT S-parameters from quasioptics, including the effects of vacuum windows.  In Sec .~\ref {sec:design1},  the system configuration is presented and evaluated with numerical simulation via physical optics (PO). In Sec.~\ref{sec:calibrat}, several system tests are performed including system throughput, vacuum windows displacement, and planar near-field scanning of the beam at the reference plane. A description of the calibration approach is then provided. In Sec.~\ref{sec:results1}, numerous DUTs are measured under room temperature and cryogenic conditions, measurements. The results are further discussed in Sec.~\ref{sec:discuss}, including the source of errors after calibration. Additionally, enhancements to the measurement approach and calibration methodology are proposed to improve sensitivity and reliability. We believe these findings will be highly relevant for testing mm-wave and THz devices in radio astronomy applications, as well as for supporting quantum computing components research at submillimeter wave and THz frequencies.
\section{\label{sec:design1}Design and simulations}

\subsection{System description and preliminary estimations}
\begin{figure*}
    \centering
    \includegraphics[width=\linewidth]{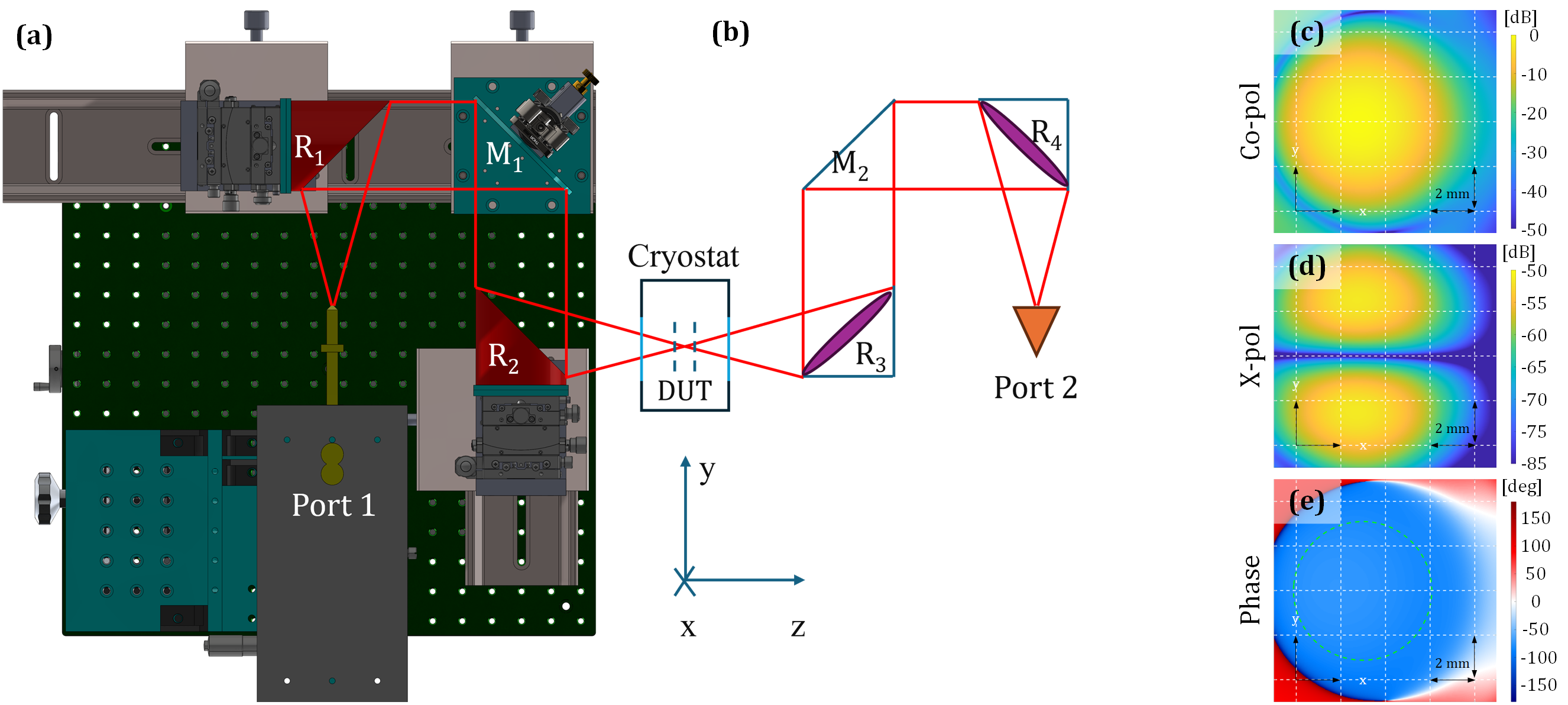}
    \caption{Design and simulation of the quasioptical setup:  CAD model of the left part of the quasioptical setup (a) with a schematic image of the beam evolution (b); and physical optics simulations of the beam inside the cryostat at $f=$ 272 GHz (desired antenna central frequency), namely (c) co-polarized beam profile at $z=0$, (d) cross-polarized beam profile at $z=0$, and (e) phase distribution within the co-polarized beam, dashed lines correspond to the beam waist level in Gaussian-beams, i.e., $1/e$ power change. Since the cryostat is located on the translation stage, one can also independently use the setup for room-temperature measurements. The distance between each quasioptical element is $d=f=152.4$ mm, except the double focal distance $2f$ as the optical (not physical) path between $R_2$ and $R_3$. The system has a mirror symmetry with respect to $x-y$ plane, and the inter-grid space is 2 mm.}
    \label{fig:Simulation2}
\end{figure*}
The quasioptical system design is represented in Fig.~\ref{fig:Simulation2} (a)-(b). The setup consists of four parabolic reflectors (focal length $f=152.4$ mm) and two flat mirrors coupled to two Tx/Rx modules with 220–330 GHz extenders connected to a VNA. The extenders are calibrated with two-port flange-to-flange calibration and then coupled to Picket-Potter horn antennas (23-dB gain at 272 GHz central frequency). The radiation is initially collimated with a parabolic reflector, then guided by a flat mirror toward the second parabolic reflector, which focuses the beam on the device under the test (DUT) reference plane. The system is symmetric with respect to the reference plane, so the same optics guide the waves to and from the second frequency extender.

While an ordinary system would focus the beam at the focal distance of the second parabolic mirror, the vacuum chamber quartz windows (thickness $t=3.175~\text{mm}$, inter window distance $d\approx ~92.71~\text{mm}$) displace the focal plane from the free space position. To determine the optimal position for the Gaussian-beam focus to be centered within the chamber, matrix calculations were performed, yielding $h\approx~154~\text{mm}$ for the Gaussian-beam focus relative to the focusing reflector, compared to $f=152.4 \text{mm}$ without the windows.

The Picket-Potter horn antenna used in the experiment can be approximated as a Gaussian-beam source. For material characterization applications, it is crucial that the beam remains effectively collimated at the DUT reference plane. Based on previous estimates\cite{Luomaniemi2016}, the smallest beam waist $w_0\approx1.72~\text{mm}$ for the antenna occurs at $248~\text{GHz}$, resulting in a confocal distance of $z_c=7.78~\text{mm}$. This confocal distance is an order of magnitude larger than the target devices, ensuring the beam remains approximately collimated across the space bordered by the DUT reference planes, which is $\approx$ 25 times smaller.

 A numerical simulation was carried out using the TICRA GRASP software to evaluate how well the setup would perform. Instead of using the usual Gaussian-beam approximation, we defined the source in GRASP using a far-field pattern. To get this pattern, we first simulated the Pickett-Potter horn antenna in CST Microwave Studio, making sure its design was optimized for a center frequency of 272 GHz, matching the antenna used in our actual measurements based on the original paper\cite{Potter}. The resulting far-field pattern from CST was then imported into GRASP for further analysis.

Numerical simulation results are presented in Fig.~\ref{fig:Simulation2} (c)-(e). As shown in Fig.~\ref{fig:Simulation2} (c), the beam axis is shifted by $1$ mm to the left. At the same time, the maximal beam intensity is $\approx$25 dB higher than the highest cross-polarization level. The beam resembles a slightly modified Gaussian-beam with an equivalent beam waist of $w_0\approx$ 3 mm. However, the beam deviates from a pure Gaussian shape, with stronger sidelobes observed on the left side of the figure. The vacuum windows of the cryostat perturb the beam, producing interference patterns after the beam passes through the windows. A portion of the radiation is also lost due to the reflections showing as standing wave patterns. However, the beam phase remains relatively flat across a cross sectional area that encapsulates the intended DUT size. The beam phase was estimated around the beam center, with deviations from the highest beam intensity point reaching $\Delta\phi\approx$ 5$^\circ$ at the edges of the beam waist. Beam asymmetry leads to more substantial phase deviations at distances greater than 3 mm from the beam center.

The numerical simulation was performed at the band’s center frequency and provided results consistent with the system's further construction goals. A more detailed analysis using near-field measurements is presented in subsequent sections.
\section{\label{sec:calibrat}Setup tests and calibration}

\subsection{Quasioptics alignment}

A flange-to-flange calibration was performed first using standard, manufacturer-provided\cite {vdiodes1020kit} waveguide calibration standards; short, open, line, and thru (SOLT). The system was visually aligned first with a 650 nm LED as a source, which was fixed in the horn antenna. This helped to ensure that a point source of visible light is focused on another antenna after passing the quasioptics. Then, the final alignment was performed with the VNA extenders, without the cryostat windows. Final alignment was optimized with respect to the maximal signal in $S_{21}$, and the achieved transmission response level was $-2~\text{dB}$ at the band central frequency. Final fine-tuning was performed in the time domain with windows installed, achieving the same optical path length for radiation from both VNA ports.

After visual alignment, we assume that the Gaussian beam focus is at the same position for both extenders, defining the reference plane for the DUT measurement. However, the reference plane should be kept at the DUT holder plane inside the cryostat. Since quasioptical elements for each port are built on separate breadboards connected to separate translation stages, as shown in Fig.~\ref{fig:Simulation2} (a), the relative positions between the quasioptical elements of each breadboard half were maintained, and each breadboard was moved along the focused optical axis as a whole. An optical distance sensor (Sick GmbH, 20 $\mu m$ resolution) used to monitor the distance between the stages, while each coordinate of the mechanical translation stage was validated with mechanical dial indicators (Mitutoyo, 25 $\mu m$ resolution). A $300-\mu\text{m}$ thick reflector was inside the cryostat as a reference, and the movement was performed with respect to the $S_{11}$ and $S_{22}$ signals in the time domain. Overlapping signal peaks indicate that the Gaussian-beam waist is located at the device holder plane for both ports. The corresponding configuration also leads to approximately the maximum value of reflection signals. The presence of a holder with $8-\text{mm}$ round aperture does not noticeably affect the measured S-parameters with respect to free space ones, thus supporting the optical simulations and measurements reported above.

\begin{figure}
    \centering
    \includegraphics[width=\linewidth]{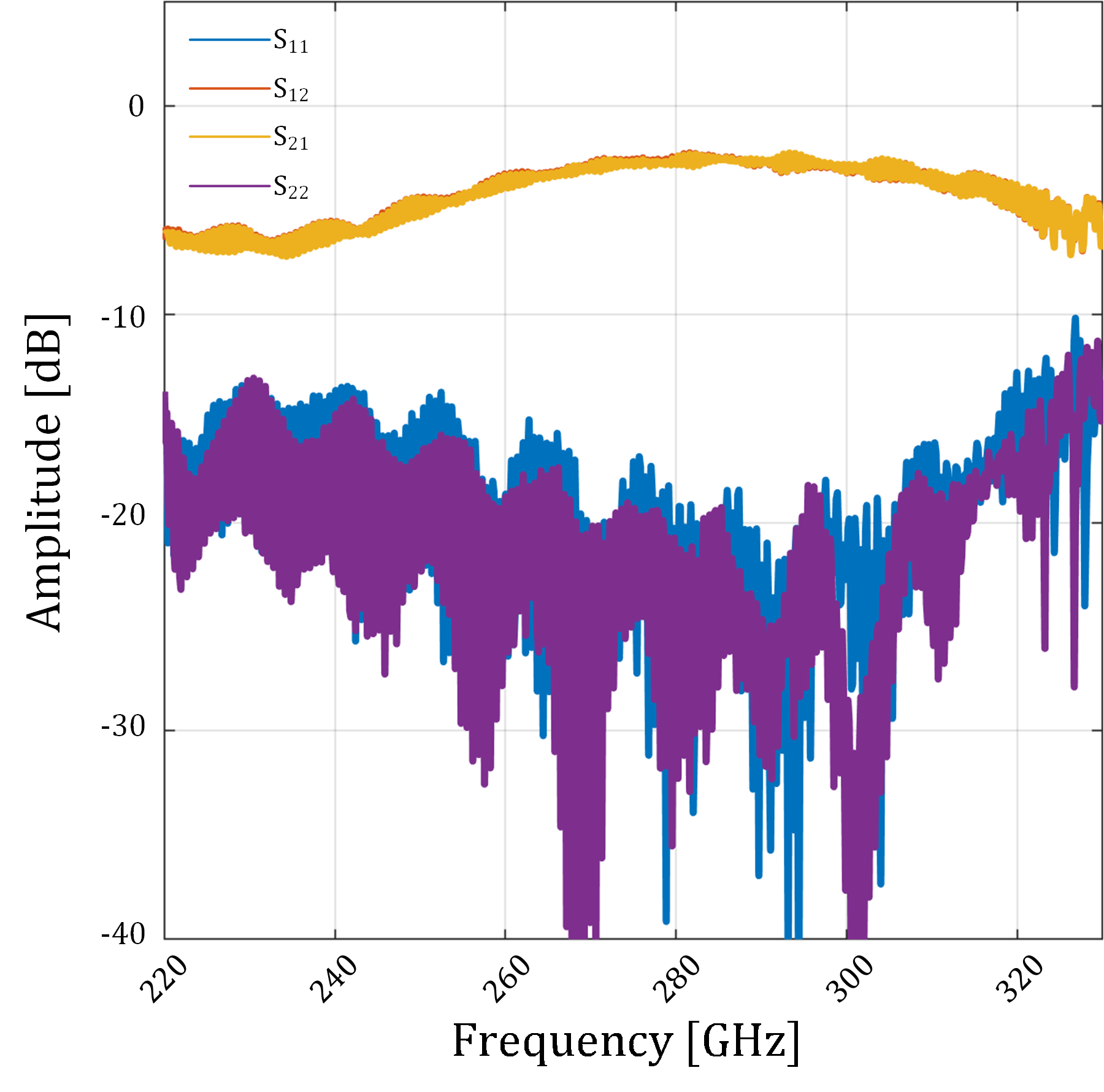}
    \caption{Measured S-parameter amplitudes of the waveguide-flange calibrated quasioptical system.}
    \label{fig:alignment}
\end{figure}

The S-parameters measured after final alignment can be seen in Fig.~\ref{fig:alignment}. We were able to reach $-2.5$ dB max transmission through the system at the mid band and $-6$ dB at the band edges. Maximum transmission/minimum return loss occurs around 290 GHz and not around the antenna's center frequency, which is 272 GHz. The return loss of the system is around $18$ dB at the midband and $12$ dB at the band edges. Mounting the cryostat windows, in turn, adds a Fabry-Perot pattern to the S-parameter curves, so it is omitted here.

\subsection{Near-field measurements}
\begin{figure}
    \centering
    \includegraphics[width=\linewidth]{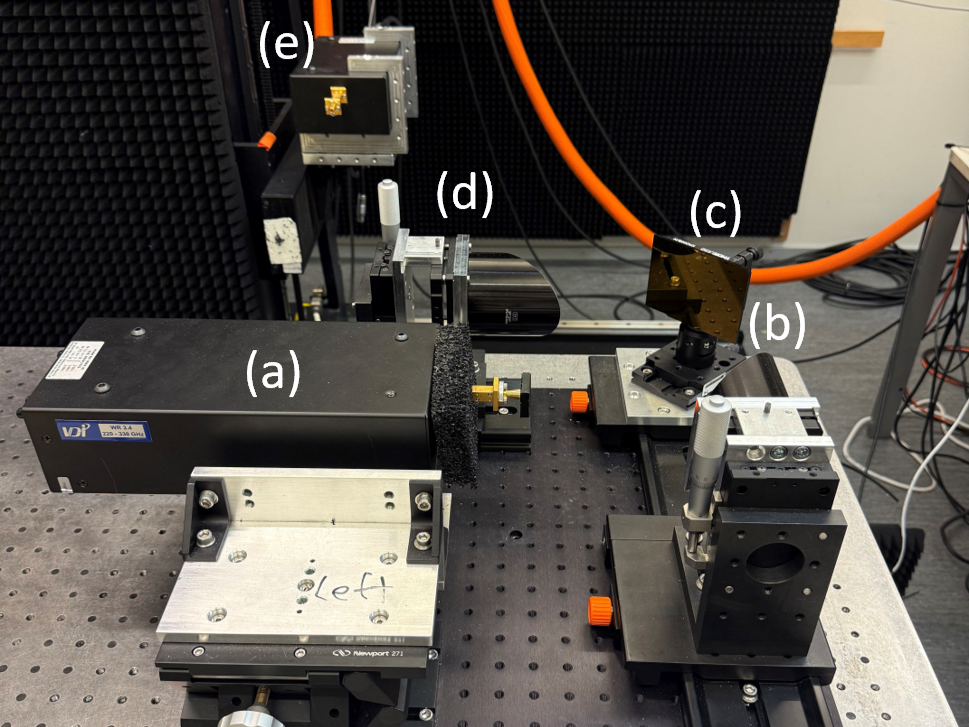}
    \caption{Near-field measurements system with a half of the quasioptical system: (a) WR 3.4 frequency extender with a Pickett-Potter horn antenna; (b) parabolic reflector $R_1$; (c) flat mirror $M_1$; (d) parabolic reflector $R_2$; (e) WR 3.4 extender with a waveguide probe.}
    \label{fig:NF_photo}
\end{figure}

Following system alignment, the left side half breadboard was selected for beam profile measurements. The breadboard with quasioptics and extenders was disconnected and mounted infront of a planar near-field scanner (NFS). The quasioptical system was fed by the same millimeter-wave extender (WR3.4-VNAX by Virginia Diodes Inc.) driven by a vector network analyzer (N5225B PNA by Keysight Technologies). Scanning in the measurement range was performed with the NSI 200V-5×5 by Near-Field Systems Inc., as shown in Fig.~\ref{fig:NF_photo}.

\begin{figure}
    \centering
    \includegraphics[width=\linewidth]{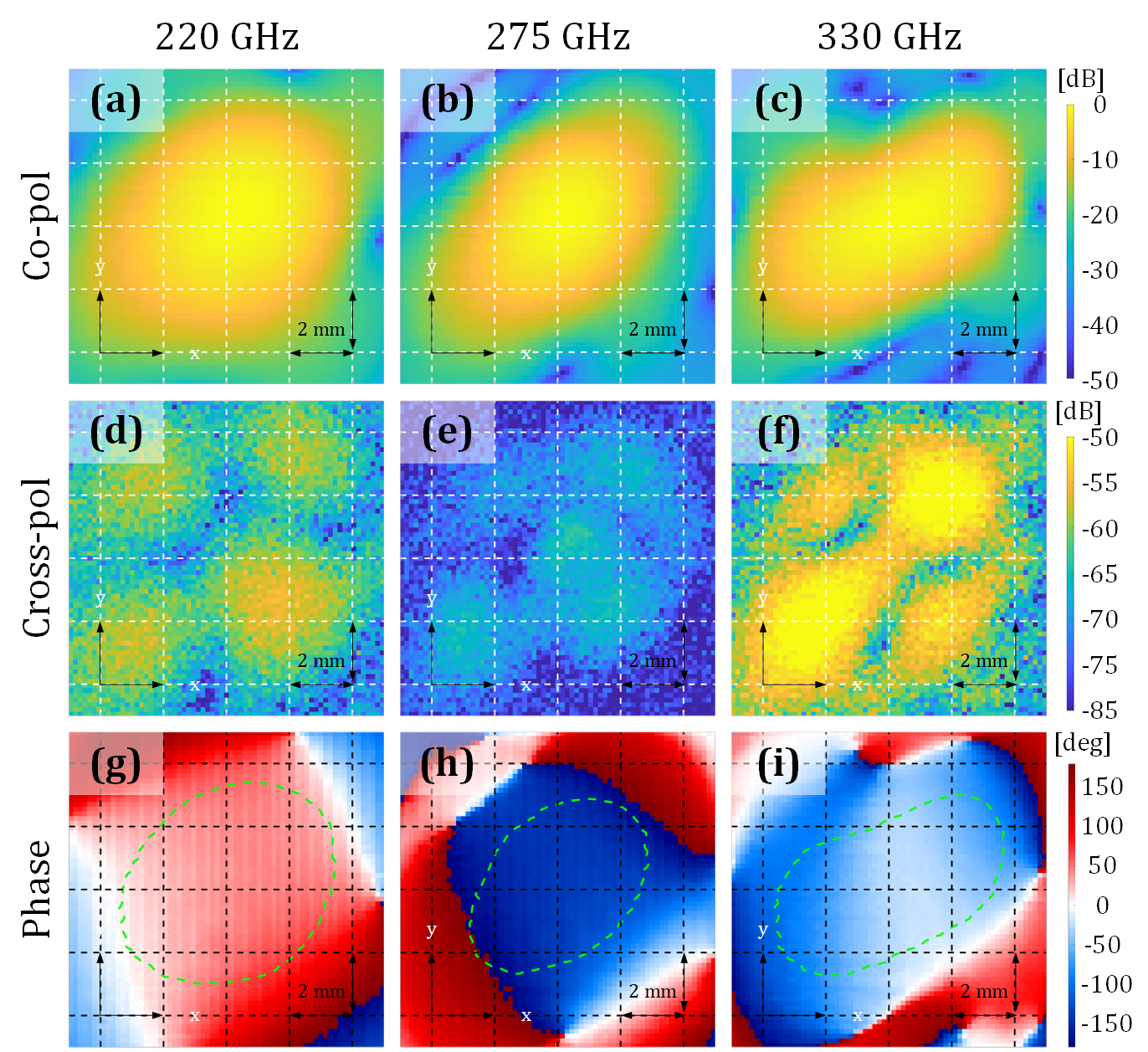}
    \caption{Near field measurements of the beam: (a) co-polarized beam pattern at 220 GHz, (b) co-polarized beam pattern at 275 GHz, (c) co-polarized beam pattern at 330 GHz, (d) cross-polarized beam pattern at 220 GHz; (e) cross-polarized beam pattern at 275 GHz, and  (f) cross-polarized beam pattern at 330 GHz. Co-polarized phase distribution patterns for (g) 220 GHz, (h) 275 GHz, and (i) 330 GHz. Dashed green lines correspond to $1/e$ power change, and the intergrid line space is 2 mm. Cross-pol is normalized with respect to the corresponding Co-pol, for convenience.}
    \label{fig:NFP}
\end{figure}

As shown by the measurement results in Fig.~\ref{fig:NFP}, the amplitude of the focused beam is different from Gaussian. Deviations from the Gaussian profile increase as a function of frequency, as misalignments are more easily visualized at shorter wavelengths, leading to more pronounced negative effects. Nevertheless, the beam at the waist is strongly linearly polarized, with the cross-polarization level averaging 50 dB lower than the co-polarized level.

One of the main concerns for further materials characterization is phase variations at the reference plane. Phase variation was also measured using the NFS, and the results are presented in Fig.~\ref{fig:NFP}. We found that for the co-polarized radiation, the phase variation is relatively small, with an average difference of 8$^\circ$ in the beam waist area at 220 GHz, 9$^\circ$ at 275 GHz, and 11$^\circ$ at 330 GHz. These values are within a satisfactory range defined in works reported before \cite{legg2013implementation,bourreau2006quasi}.

\subsection{Quasioptical calibration}

\begin{figure}
    \centering
    \includegraphics[width=\linewidth]{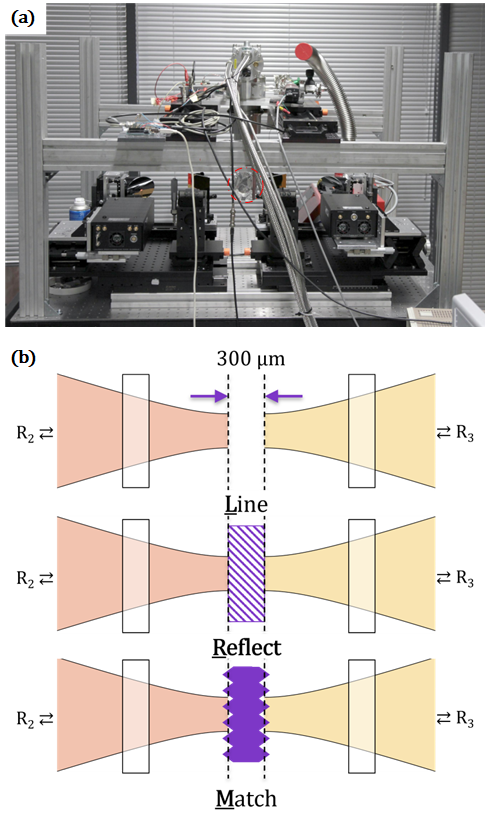}
    \caption{(a) Photo of the system with the cryostat (marked in red circle) and (b) Line-Relect-Line (LRM) calibration procedure (Not to scale): firstly, a 300 $\mu m$ line is measured, then a reflection standard of the same thickness, and, finally, a double-sided match. $R_2$ and $R_3$ are parabolic reflectors with respect to Fig.~\ref{fig:Simulation2}.}
    \label{fig:calibration}
\end{figure}

After waveguide calibration, the second-tier calibration removes the impact of the quasi-optics and establishes the measurement's reference plane at the surface of the sample holder. The sample holder was specifically designed to securely position both the samples and the calibration standards, as the precise alignment of the reflection calibration standard determines the reference plane for all measurements.

Mathematically, the calibration mechanism can be explained with ABCD matrices. The real DUT ABCD matrix is embedded in an error network, described as\cite{cataldo2015analysis}:

\begin{equation}
\boldsymbol{DUT_m}=\boldsymbol{X*DUT*Y},
\end{equation}
where $\boldsymbol{DUT_m}$ is the signal measured by ports, $\boldsymbol{X}$ and $\boldsymbol{Y}$ is a fixture (error) network, which includes impact from vacuum windows, device fixtures, etc. When calibration standards are measured inside the cryostat, their real and measured S-parameters are used to find $\boldsymbol{X}^{-1}$ and $\boldsymbol{Y}^{-1}$ so the real DUT S-parameters can be simply found as  $\boldsymbol{DUT}=\boldsymbol{X}^{-1}\boldsymbol{DUT_m}\boldsymbol{Y}^{-1}$.

Due to the presence of the optical windows, the LRM calibration method was utilized (Fig. \ref{fig:calibration}). The cryostat has four optical windows: two are used to direct radiation to the DUT, while the other two allow the DUT to be mounted or replaced without interfering with the optical path. For the line standard, the holder was left empty, and a delay corresponding to $L=300~ \mu\text{m}$ was introduced into the calibration program. This delay was chosen to prevent incorrect phase retrieval during calibration by ensuring the relationship $10^\circ < k \cdot L < 170^\circ$ was satisfied, where $k$ is the wavenumber in free space. A $300-\mu \text{m}$ thick stainless steel plate was used as the reflection standard. Pyramidal absorbers, TK-RAM (Thomas Keating Ltd, UK)\cite{thomas}, were used as match standards.  The calibration gives approximately $30~\text{dB}$ over the band of return loss when the cryostat is at vacuum and empty.

\begin{figure}
    \centering
    \includegraphics[width=\linewidth]{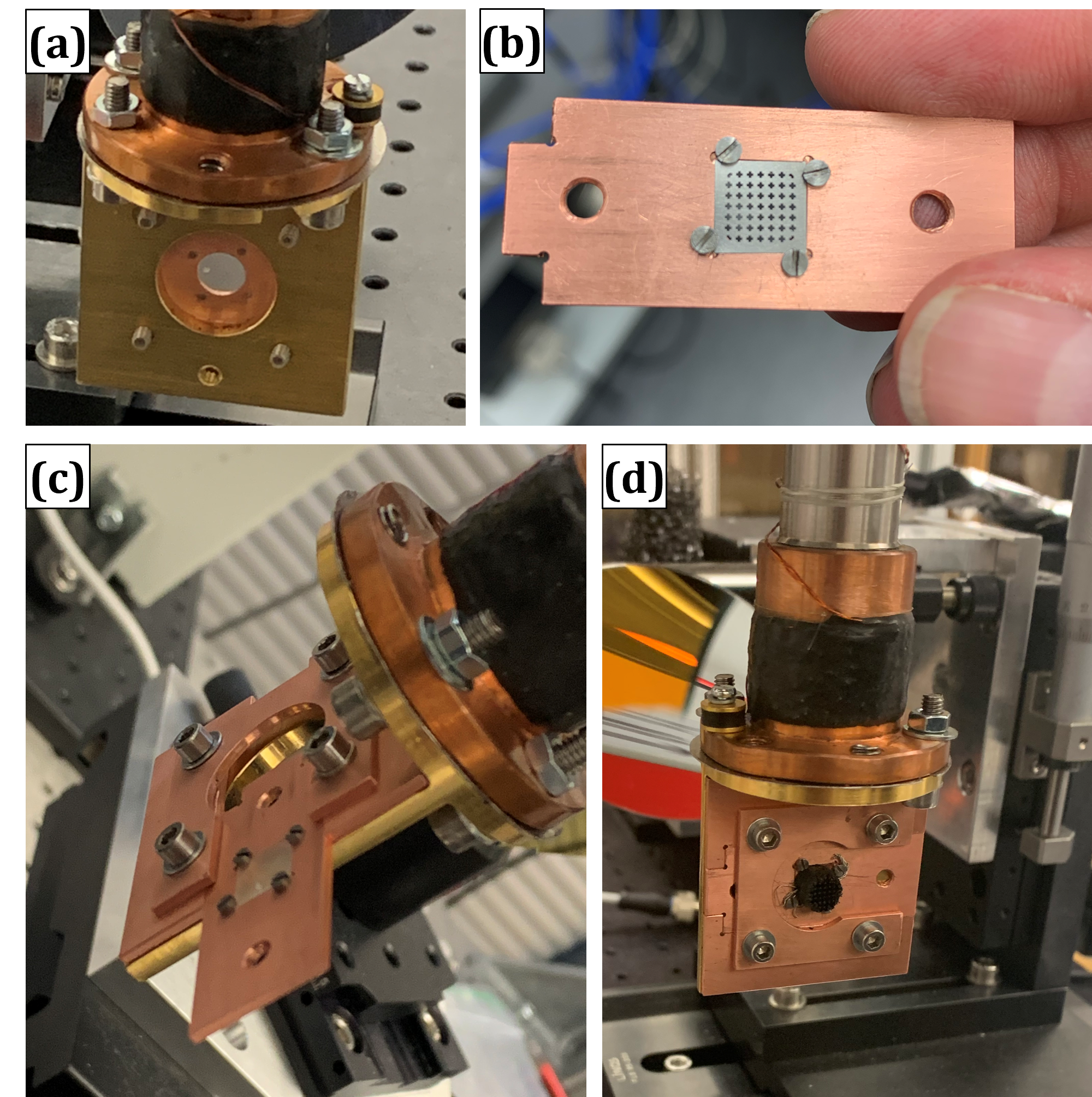}
    \caption{Device fixture and the cold finger: (a) cold finger aperture, (b) DUT sliding mount, (c) device installation inside the cold finger fixture and (d) an example of the DUT (Match standard) sitting on device fixture.}
    \label{fig:fixture}
\end{figure}

The DUTs (or calibration standards) have been fixed with four screws in a copper holder inside the Janis SHI 4-1 cryostat equipped with Pfeiffer Vacuum pumping system, and the pressure has been measured with a vacuum gauge from the same company.  The shield-free device fixture photo is represented in Fig. \ref{fig:fixture}. The device fixture is accessed through side optical windows that are not in the optical path (not shown). This approach results in unperturbed optical window locations and ensures the reliability and applicability of the measurements and, more importantly, the consistency of the calibration. After the device has been placed in the cryostat, air is pumped out, and the cryostat is turned on, cooling the system in general. To ensure better thermal contact, the device sliding fixture has been lubricated with a cryogenic vacuum grease gel.

During the calibration process, it was observed that the optical windows are shifted when the pumping system is activated. As a result, calibration must be performed only after a desired pressure level (in our case, approximately around 1 Pa) is reached and the positions of the sealing O-rings and windows have stabilized. Window displacement as a function of pressure was measured with an optical distance sensor. The results show that calibration cannot be performed without first pumping the vacuum in the cryostat, as optical window displacement occurs immediately after the pump is switched on. Specifically, vacuum conditions cause the windows to shift by approximately 130 $\mu \text{m}$. This value is approximately 10 \% of the shortest wavelength, and definitely distorts the calibration, while after the calibration, one can measure the distance with 20 $\mu m$ accuracy, which is 2 \% from the shortest wavelength, which we consider to be accurate for our measurements.

\begin{figure}
    \centering
    \includegraphics[width=\linewidth]{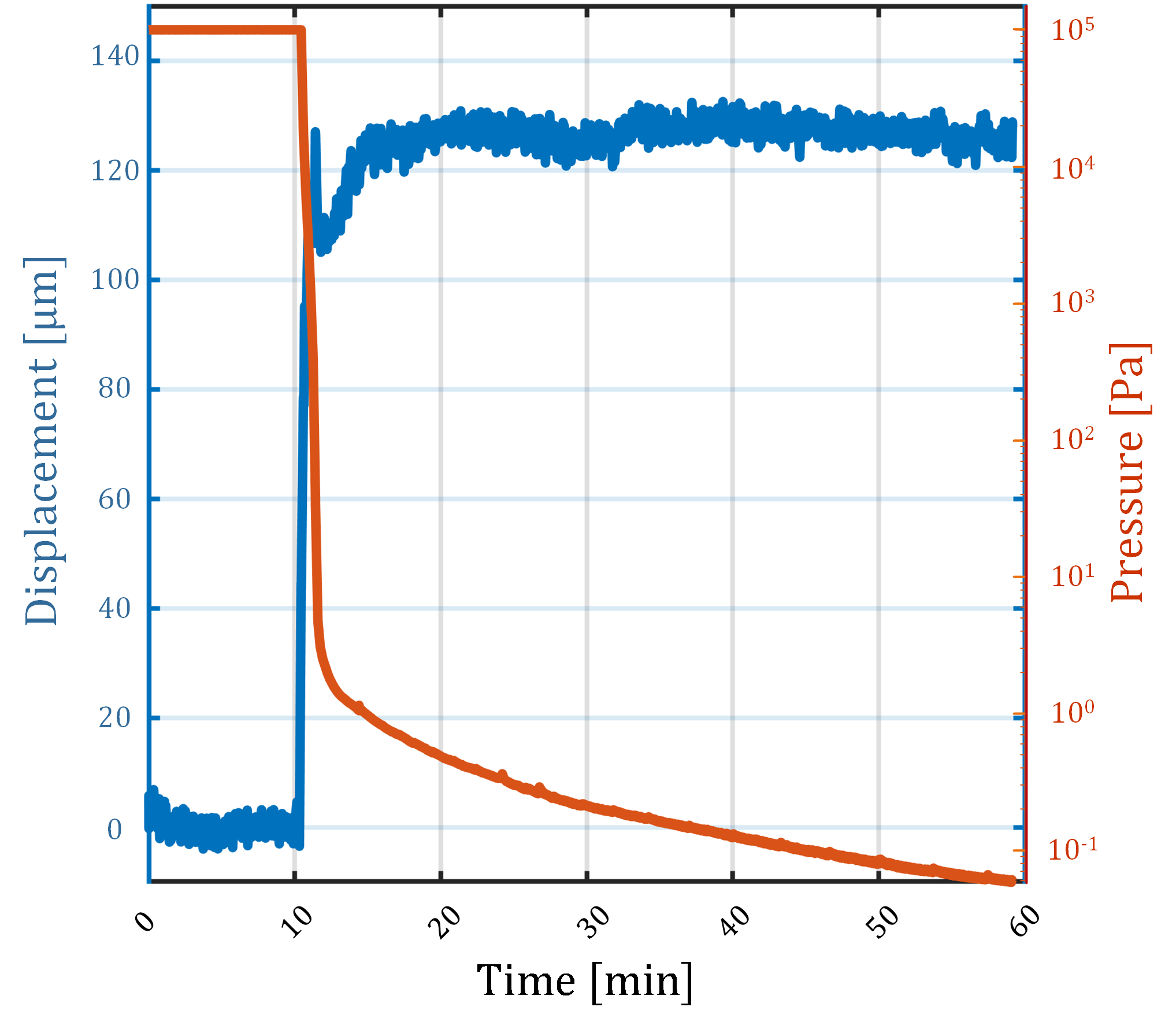}
    \caption{Vacuum window displacement due to changing pressure. The pump was switched on after 10 minutes of measurements.}
    \label{fig:Pressure}
\end{figure}

The measurements indicate that the calibration can be performed once the system stabilizes, which typically occurs after about 10 minutes of pumping, according to Fig.~\ref{fig:Pressure}. Significant pressure changes occur during the first 5 minutes of pumping, leading to the most noticeable window displacement. The subsequent 5 minutes result in smaller pressure changes, allowing the vacuum windows to stabilize, so we can perform the calibration around 1 Pa pressure. Once the system reaches stability, measurements for calibration calculations can begin, as the device under test (DUT) is measured under high-vacuum conditions.

\begin{figure}
    \centering
    \includegraphics[width=\linewidth]{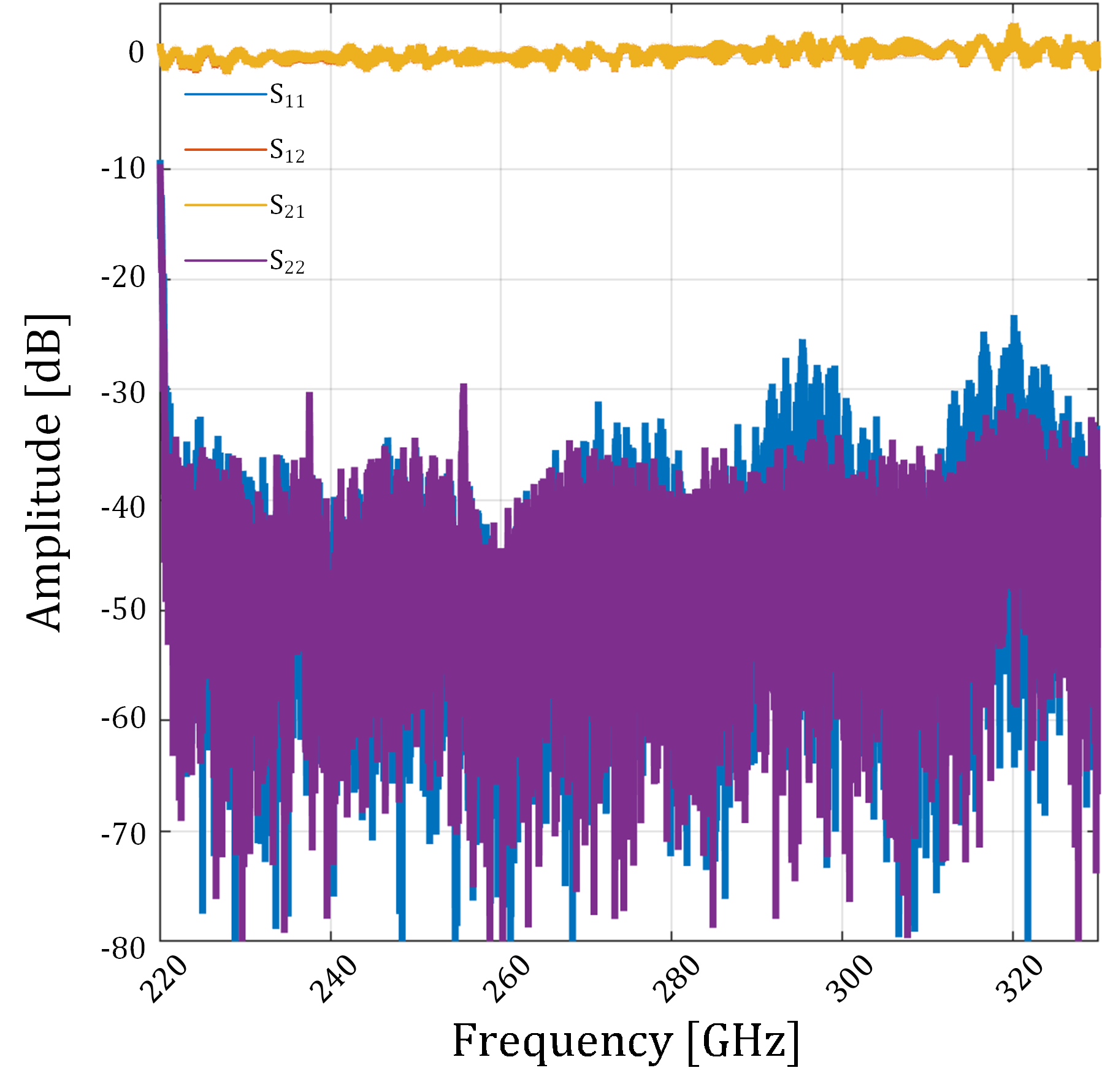}
    \caption{Validation of the quasioptical calibration via measurements of the empty holder, i.e. line standard.}
    \label{fig:calvalidation}
\end{figure}

The calibration results were validated by measuring the empty holder after the calibration correction was applied. In this case, the line standard was measured last, and the S-parameters, immediately following calibration are reported in Fig. \ref{fig:calvalidation}.

The return loss in the sample-free system is approximately 30 dB, but tends to decrease over many pumping cycles. Experimental evidence suggests a new LRM calibration should be performed before each DUT measurement.
\section{\label{sec:results1}Results}
\subsection{Device simulation and fabrication}

\begin{figure}
    \centering
    \includegraphics[width=\linewidth]{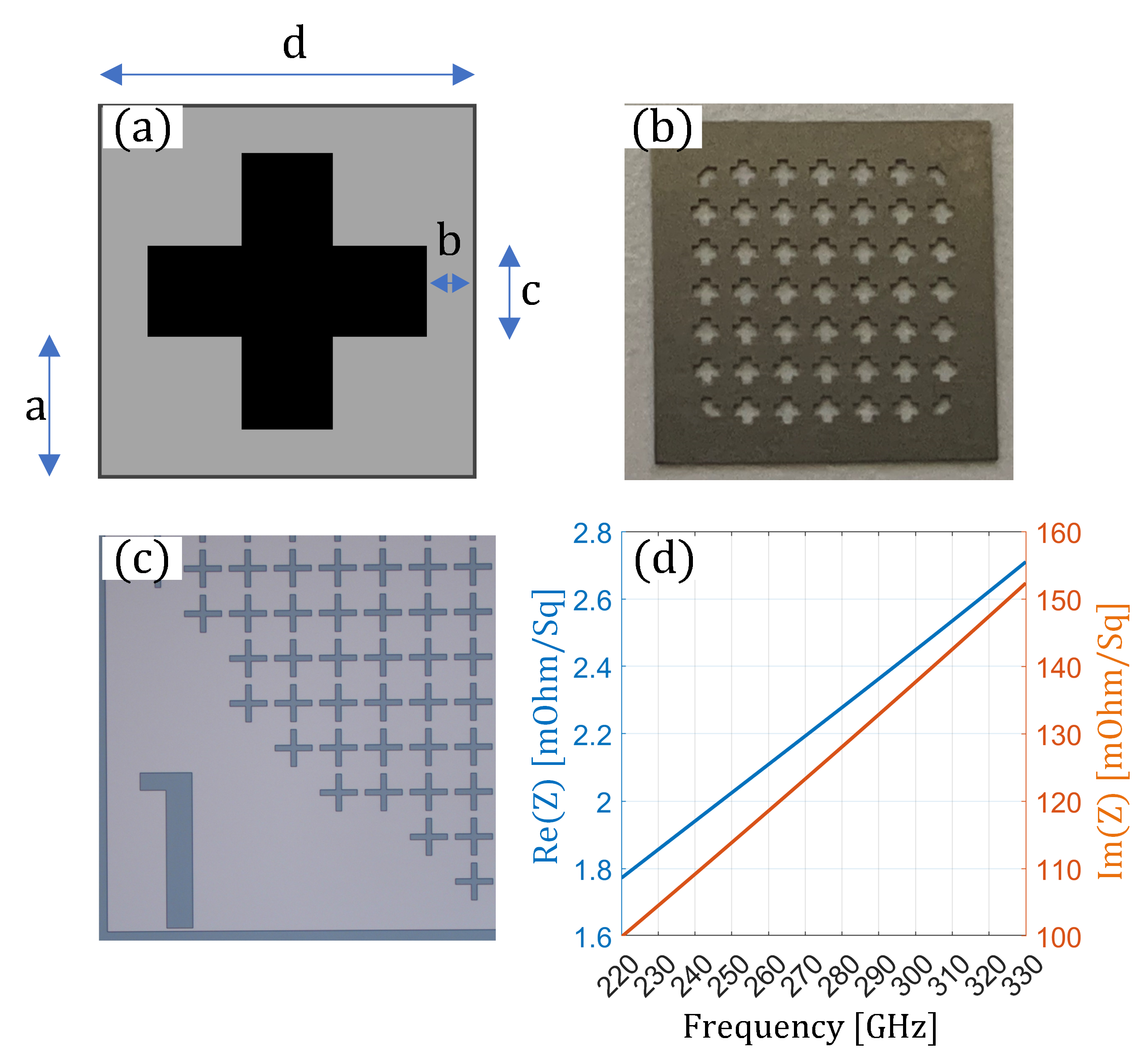}
    \caption{DUTs parameters used in the study: (a) design of the FSS unit cell; (b) photograph of stainless steel filter; (c) microscope view of a niobium-based FSS; (d) ac surface impedance (Z) of Nb films at T=4.8 K calculated with MB approach.}
    \label{fig:DUTS}
\end{figure}

Three DUTs were selected for characterization. The first DUT is a $t=280 ~\mu m$ high-resistivity silicon wafer, and its theoretical performance can be estimated based on previously obtained data\cite{buechler1986silicon,lamb1996miscellaneous}. The next DUTs are described in Fig.~$\ref{fig:DUTS}$, namely, cross-shaped bandpass filters, common devices in millimeter-wave astronomy instrumentation\cite{perido2022cross,porterfield1994resonant}. The first filter under test is a classic design \cite{porterfield1994resonant} and, in this case, fabricated from stainless steel (Fig.~$\ref{fig:DUTS}$ (b)). The second one is an analogous cross-shaped niobium-on-silicone FSS (Fig.~$\ref{fig:DUTS}$ (c)) with temperature-dependent characteristics. The stainless steel DUT unit cell has the following dimensions: square patch size $a=365~\mu \text{m}$, cross-to-edge distance $b=185~\mu m$, cross thickness $c=270~ \mu \text{m}$, and the unit cell periodicity $d=1000 ~\mu \text{m}$ with stainless steel thickness of $t=100~\mu \text{m}$. The niobium-based filters have the following dimensions: $a=114~\mu \text{m}$, $b=23 ~\mu \text{m}$, $c= 32 ~\mu m$ and $d=260 ~\mu \text{m}$, being fabricated on $t=350 ~\mu\text{m}$ high-resistivity silicone wafer ($\epsilon=11.49$ at cryogenic temperatures\cite{wafer}). The stainless steel FSS was fabricated using laser cutting technology, while the niobium-based FSSs were fabricated via lithography and electron beam evaporation.  At the same time, the niobium layer conductivity, which transitions to it's superconducting state below the critical temperature $T_c$, was estimated using Mattis-Bardeen (MB) theory\cite{Zmuidzinas} with $T_c=9.05~K$, superconducting energy gap ($\Delta_0$) ratio $\frac{\Delta_0}{k_B T_c}=1.76$ and room temperature resistivity of $\rho=24~\mu \Omega \cdot$cm. MB theory utilizes dc parameters to calculate the superconducting conductivity of a material by using its normal resistivity and critical temperature $T_c$ values. The calculated conductivity of Nb film at 4.8 K was converted to surface impedance represented in Fig.~$\ref{fig:DUTS}$ (c) for further numerical simulations. CST Microwave Studio suite was employed to simulate the DUT spectral properties. In both cases, unit cell boundary conditions were applied in the frequency domain solver. While Nb-based DUTs are not the most suited for the WR 3.4 band applications (as compared to e.g., MgB$_2$ due to lower critical temperature) their relatively weak tunability at low temperatures provides a challenging test of system sensitivity to small changes in DUT performance. 
\subsection{Measurements}

\begin{figure}
    \centering
     \includegraphics[width=\linewidth]{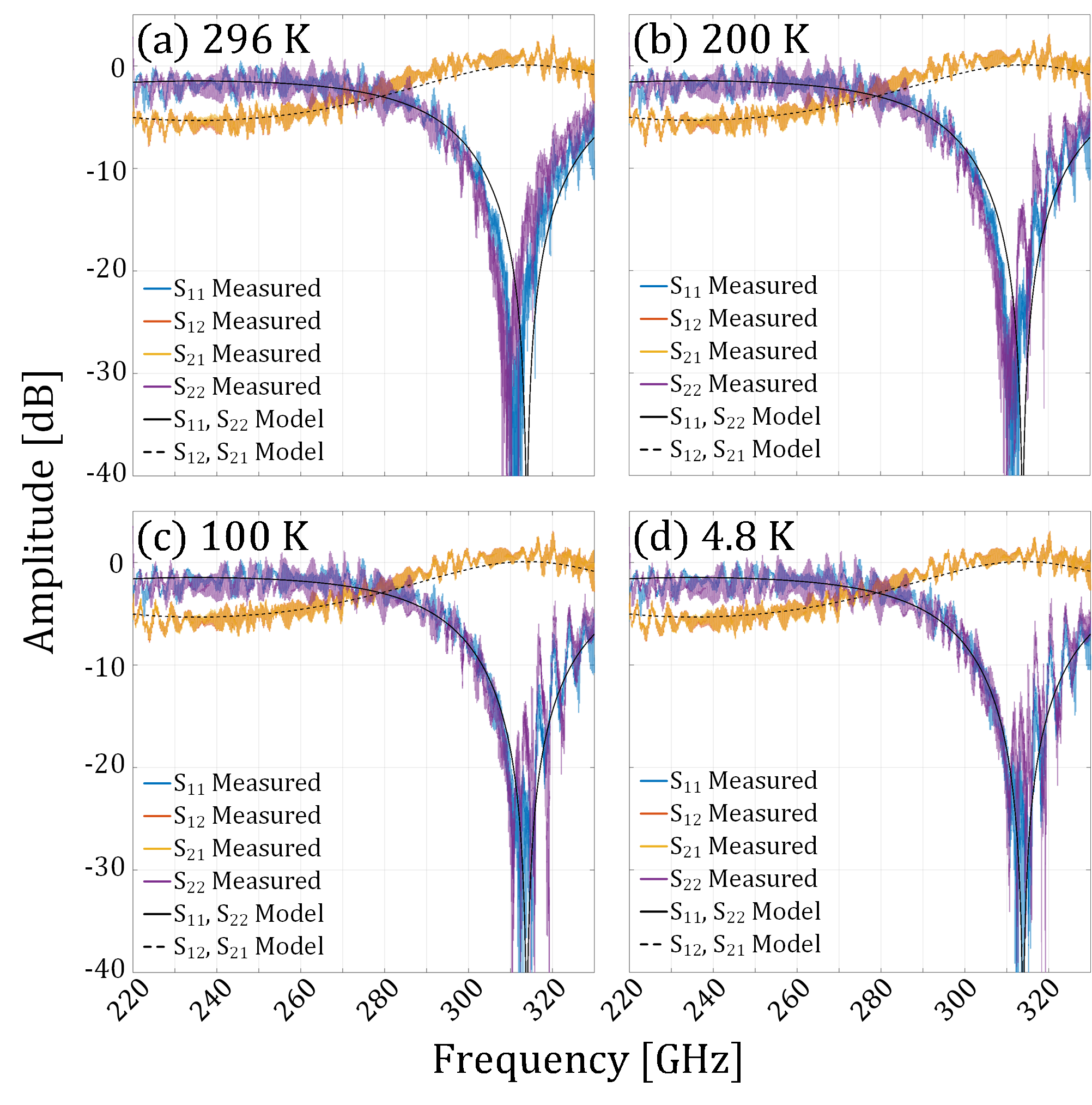}
    \caption{S-parameters of a silicon wafer: (a) 296 K (room temperature); (b) 200 K; (c) 100 K; (d) 4.8 K. S-parameters of the silicon wafer were calculated according to Ref. \cite{wafer} and only low-temperature permittivity value has been used, to which the S-parameters tend to shift as the sample is cooled down.}
    \label{fig:Si}
\end{figure}

First, the performance of the system was tested with a $240 ~\mu m$ silicon wafer. The S-parameters of the wafer have also been simulated with Fresnel formulas. The results are presented in Fig.~\ref{fig:Si}, measured at four temperature points: room temperature $T=296~\text{K}$, $T=200~\text{K}$, $T=100~\text{K}$, $T=4.8~\text{K}$. DT-670 silicon diode is used as a thermometer. As we can see from Fig.~\ref{fig:Si}, the measurements are quite consistent with the numerical simulation, and the resonance appears around 315 GHz, as expected.  The initial permittivity of the high-resistivity silicon wafer has not been evaluated, but as the system is cooled down, it tends to correspond with the low-temperature high-resistivity silicon value of  $\epsilon=11.49$\cite{wafer}. As temperature decreases, the relative permittivity of silicon decreases, thus leading to a blue shift in resonance. This effect might be attributed to different physical scenarios, such as reduced lattice vibrations, electron-phonon interactions, or lower polarizability\cite{exp1,exp2}.

\begin{figure}
    \centering
     \includegraphics[width=\linewidth]{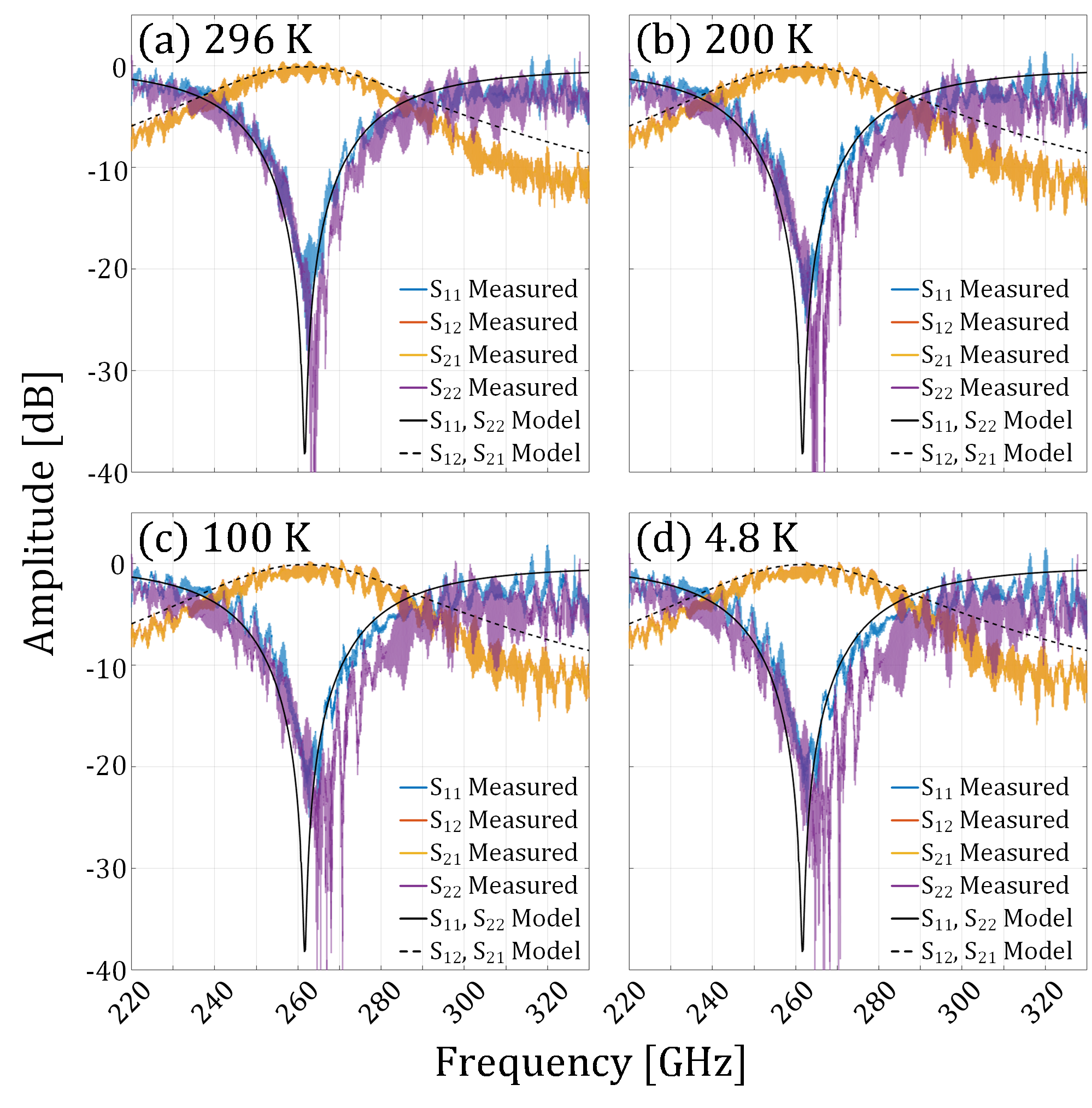}
    \caption{S-parameters of a stainless steel bandpass filter: (a) 296 K (room temperature); (b) 200 K; (c) 100 K; (d) 4.8 K.}
    \label{fig:SS}
\end{figure}

Next, we examine the behavior of the stainless steel filter. We also see that the results are similar to the numerical simulation of the stainless steel filter in Fig.~\ref{fig:SS}. The difference between numerical simulations and the experiment is likely due to fabrication inaccuracies, which are most pronounced in the corners of the plus-shaped apertures, which are rounded. Additionally, the unit cell numerical simulation assumes an infinite lateral extent device composed of the literature-based conductivity of $\sigma=7.69\cdot10^6$ S/m. The dc conductivity of our finite filter is assumed to be the literature reported value. 

\begin{figure}
    \centering
     \includegraphics[width=\linewidth]{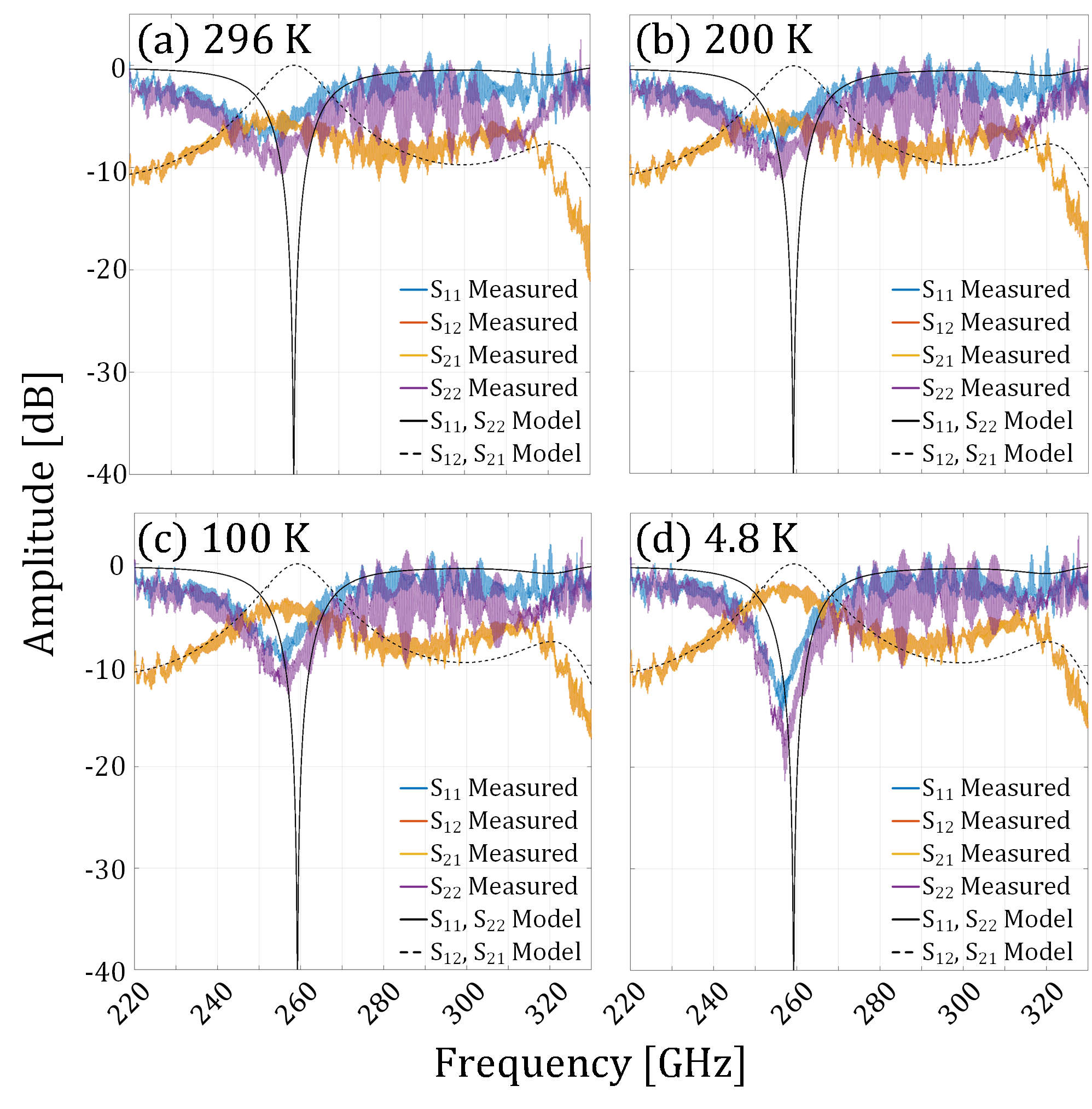}
    \caption{S-parameters of a superconducting filter prototype: (a) 296 K (room temperature); (b) 200 K; (c) 100 K; (d) 4.8 K. Numerical results are provided for T=4.8 K}
    \label{fig:SC}
\end{figure}

The final step was the S-parameter measurement of superconducting Nb filters. The results are shown in  Fig.~$\ref{fig:SC}$. The resonance peak corresponds well to numerical simulations. However, the performance of the measured filters is worse than that of the simulated ones; namely, the depths in return loss at the resonant frequency are not as sharp as in the simulation. It was reported that the Mattis-Bardeen integral, based on dc values of resistivity and critical temperature, might poorly predict the thin film behavior due to uncertainties of normal state resistivity along the wafer or superconducting gap parameter, and real values are predicted from measuring resonant devices, such as transmission line resonators \cite{tan2024operation}. Moreover, the thermometer is mounted inside the device holder and may warm up the device slightly, leading to the actual temperature of the device being higher than it is shown. Also, the absence of infrared filters may let infrared photons break Cooper pairs as well. These aspects are left for further setup modification and research. Nevertheless, both transmission and reflection S-parameters show similar trends with respect to the simulation; however, better approaches for modelling superconducting devices, for example, based on AC-extracted kinetic inductance from high-Q resonators measurements, should be implemented at these frequencies.

\section{\label{sec:discuss}Discussion}

The calibration performed at room temperature and vacuum produced consistent results, enabling the de-embedding of the DUT response from all preceding optical components. However, upon cooling the system, ripples began to appear in the S-parameters, particularly at higher frequencies. We attribute these ripples to the impact of the cryo-cooler, namely, the slow drift of the vacuum windows and the device holder, as they are less pronounced at room temperature, as shown in Fig.~\ref{fig:Si}(a), which was recorded immediately after calibration. When the device holder is cooled, minor redistribution of standing wave patterns may occur between the vacuum windows and the DUT itself. 

\begin{figure}
    \centering
    \includegraphics[width=\linewidth]{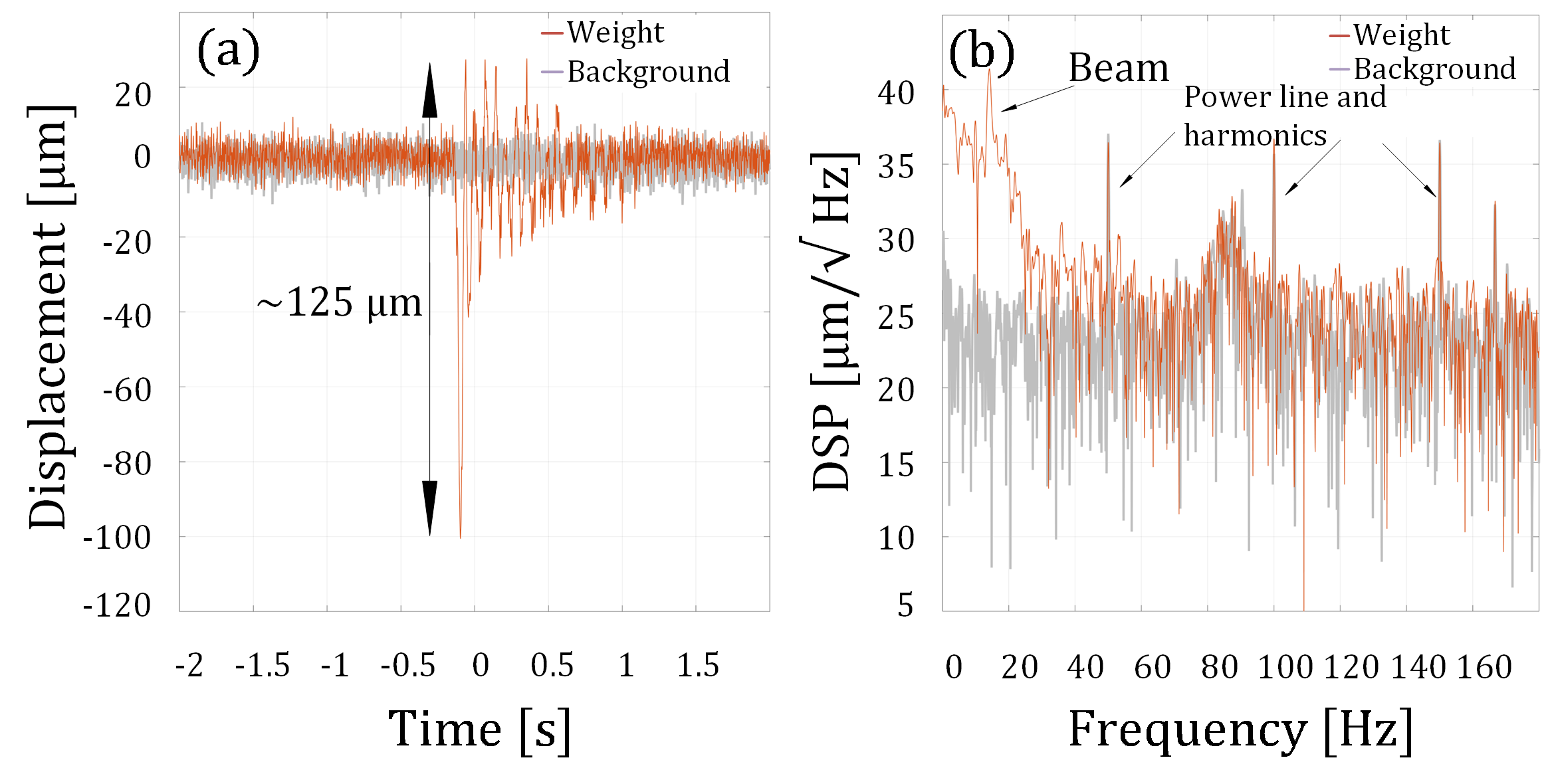}
    \caption{A displacement caused by a weight of 730 g fallen from 500 on the optical table: (a) DUT reference plane variation due to the table vibrations; (b) Displacement spectral density (DSP) of the vibrations time measurement.}
    \label{fig:displacement}
\end{figure}

Sample holder displacement was assessed by tracking its location with an optical distance sensor (Sick OD1-B100H50U25) to see if the cold finger’s flexible beam mode was contributing to sample vibration. The sensor is based on a linear CMOS array whose data is digitized at 20 bits/sample for communication via RS-232 at a maximum rate of 50 samples/second. This rate is insufficient, so the sensor analog out option was utilized, which sends the data to a 12-bit D/A converter at a rate of 1000 samples/second, and this signal was sampled with a USB-based DAC at 2000 samples/s, 14 bits/sample. The sampled signal under no externally applied disturbance is represented by the grey background trace in Fig. \ref{fig:displacement}(a). Analysis of the noise floor yields a noise limited displacement uncertainty of $\pm$ 10-15 $\mu m$.
An external disturbance was introduced as a 730 g weight dropped on the optics table from a height of 500 mm. The sample holder deflected away from the sensor 125 $\ mum$ and the subsequent damped ringing decayed to below the noise floor after 1 second, as shown by the orange trace in Figure Fig. \ref{fig:displacement} (a). The Fourier transform of both traces are shown in Fig. \ref{fig:displacement} (b) with trace colors corresponding to those in Fig. \ref{fig:displacement} (a). The flexible beam mode is quite visible at 13 Hz and the steady state sample holder location appear to be different before and after the weight drop giving rise to the 1/f-like noise visible in the image. Conversely, the flexible beam mode is not visible in the non-disturbed case. Both measurements clearly show powerline harmonics (50, 100, and 150 Hz) as well as signal from an unknown source around 80 Hz.
Overall, the dominant vibration modes of the cold finger are not present under normal operation and any subsequent higher order vibration are likely less than $\pm$ 15 $\mu m$.

First of all, we compared the vibrations caused by the cryopump by understanding the displacement caused by striking the breadboard with a 1 Kg weight to eliminate the impact of the background vibration on the device measurements. The result is represented in Fig.~\ref{fig:displacement}. As we can see, background vibrations of the system cannot be designated from the noise, so we assume that the background vibrations do not have any effect on the S-parameters.

\begin{figure}
    \centering
    \includegraphics[width=\linewidth]{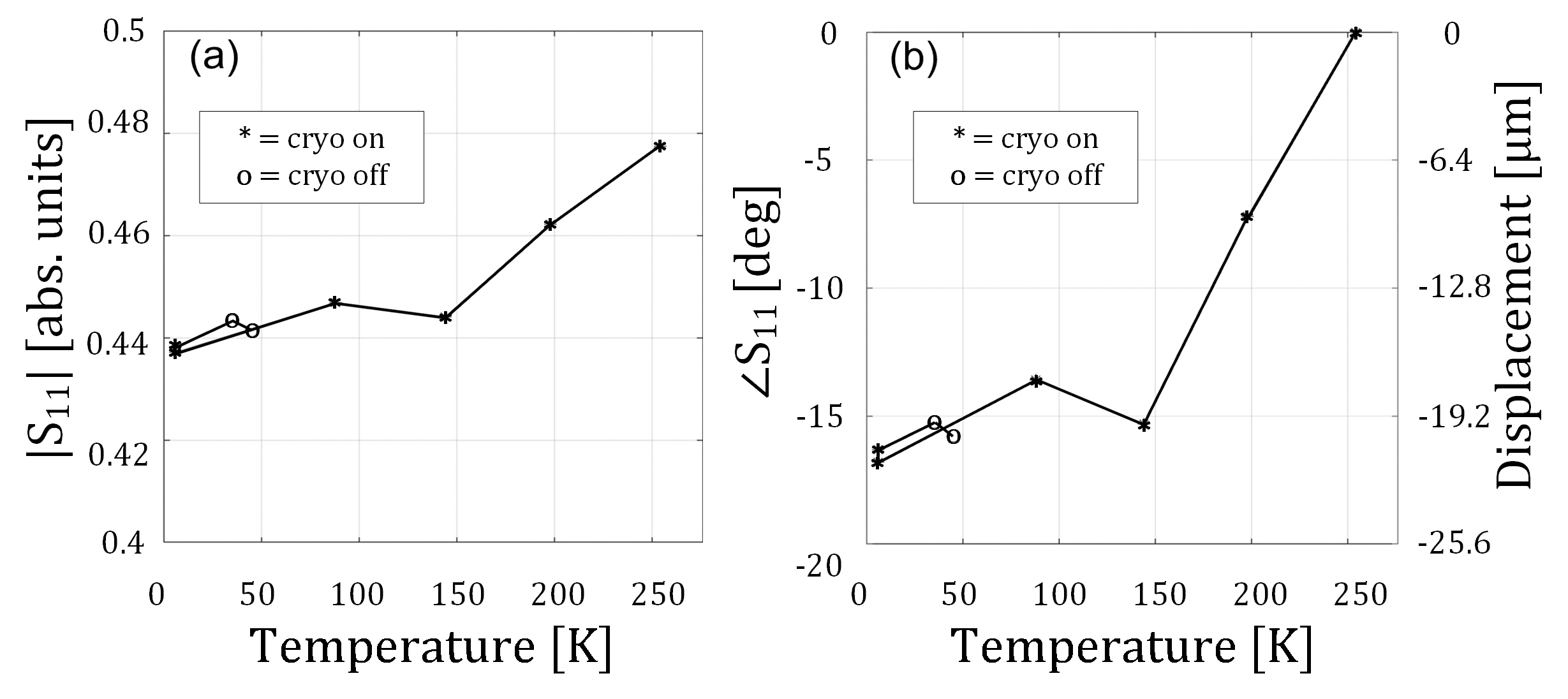}
    \caption{Stability measurement of $S_{11}$ at the band edge, i.e., 325 GHz: (a) Amplitude dependence on the temperature, (b) Phase and displacement dependence on the temperature.}
    \label{fig:stabil}
\end{figure}

Long term stability was assessed by monitoring $S_{11}$ amplitude and phase at 325 GHz (upper band edge) as the cold finger temperature was reduced from approximately 270 K to 4.8 K. The beam was transmitted through the quartz vacuum window and aligned to the side of the sample holder.  The magnitude and phase are displayed in Fig.~\ref{fig:stabil} (a) and (b), respectively. The amplitude showed a overall drop of 2 $\%$ while the phase decreased $\approx$monotonically 17 degrees from the reference point. This behaviour is consistent with a deflection of 20 microns leading to a two-pass pathlength increase of 40  microns. Such a path length change directly results in a significant phase shift, as the round-trip distance for the reflected wave changes, and the observed phase change aligns with the expected effect for this displacement, which can be represented as a load connected to a transmission line with a variable length\cite{pozar2021microwave}.

\begin{figure}
    \centering
    \includegraphics[width=\linewidth]{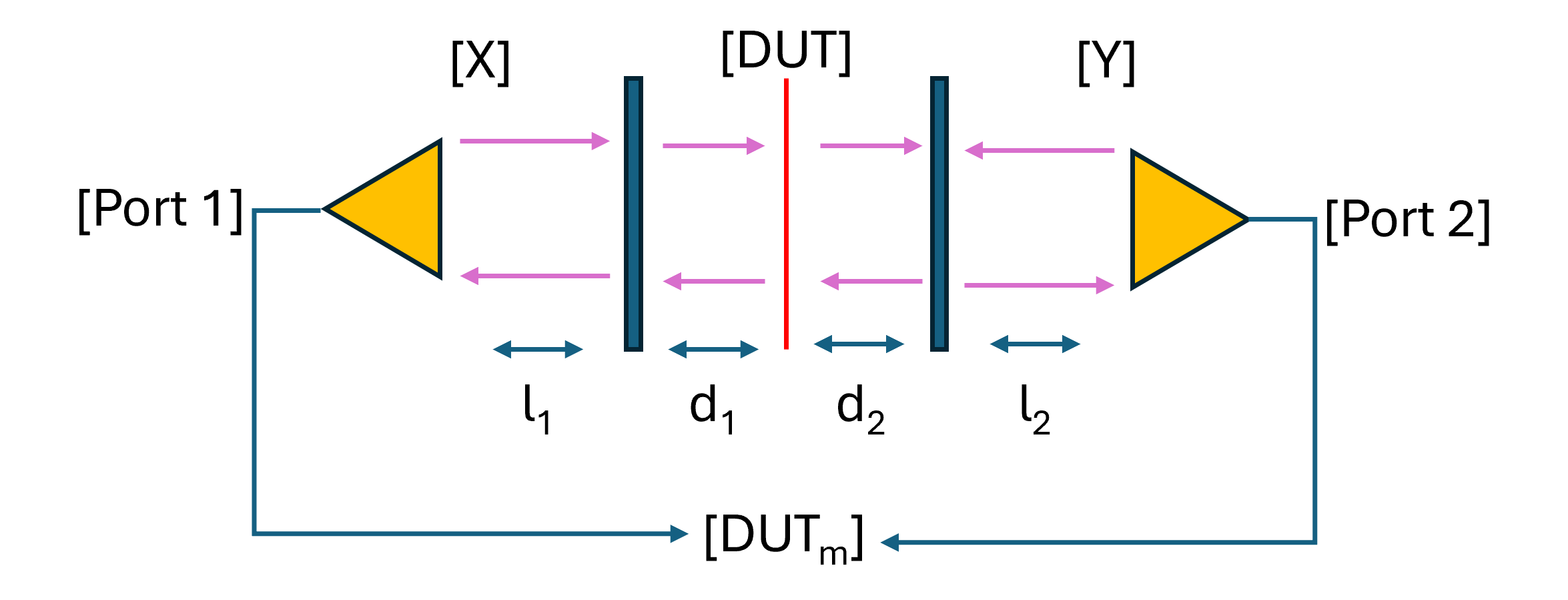}
    \caption{A simplified schematic for VNA measurements of DUT in a cryostat for numerical error analysis: $l_1=l_2=75$~mm, $d_1=d_2=49.7$~mm. Windows (blue thick lines) are taken with thickness $t=3.2$ mm (not shown) and refractive index $n=2.13$.}
    \label{fig:explanation}
\end{figure}

Different errors arising from variable error networks can be mathematically explained in this scenario with a simplified 1D system shown in Fig.~\ref{fig:explanation}. In contrast with our system, the proposed model explains the principle of quasioptical calibration with a transmission line approach. Moreover, we will restrict ourselves to the de-embedding of the DUT S-parameters from the measurement environment, i.e., we will not explicitly take into account errors like crosstalk, leakage paths, etc.

DUT in a device holder inside the cryostat, which, after calibration, is coupled to its error (fixture) network as $\boldsymbol{DUT_m}=\boldsymbol{X*DUT*Y}$ via ABCD matrix. The calibration is performed at room temperature. However, when the cryostat is switched on, some processes change the initial X-Y error network, and it becomes X'-Y':
\begin{equation}
\boldsymbol{DUT_m}=\boldsymbol{X'*DUT*Y'},
\end{equation}

 In this scenario, the de-embedded S-parameters equivalent ABCD matrix are:

\begin{equation}
\boldsymbol{DUT_d}=\boldsymbol{X^{-1}*X'*DUT*Y'*Y^{-1}}.
\end{equation}

In an ideal scenario, the products should be equal to the unity matrix

\begin{equation}
\boldsymbol{X^{-1}*X'}=\boldsymbol{Y'*Y^{-1}}=\boldsymbol{I}.
\end{equation}

However, if the environment changes, the exact condition is broken, and an error network modifies the DUT's S-parameters. Thus, results obtained further will be compromised also by different error terms, which are also changed during cooling down the system, changing the pre-calculated error network obtained with calibration.
\begin{figure}
    \centering
     \includegraphics[width=\linewidth]{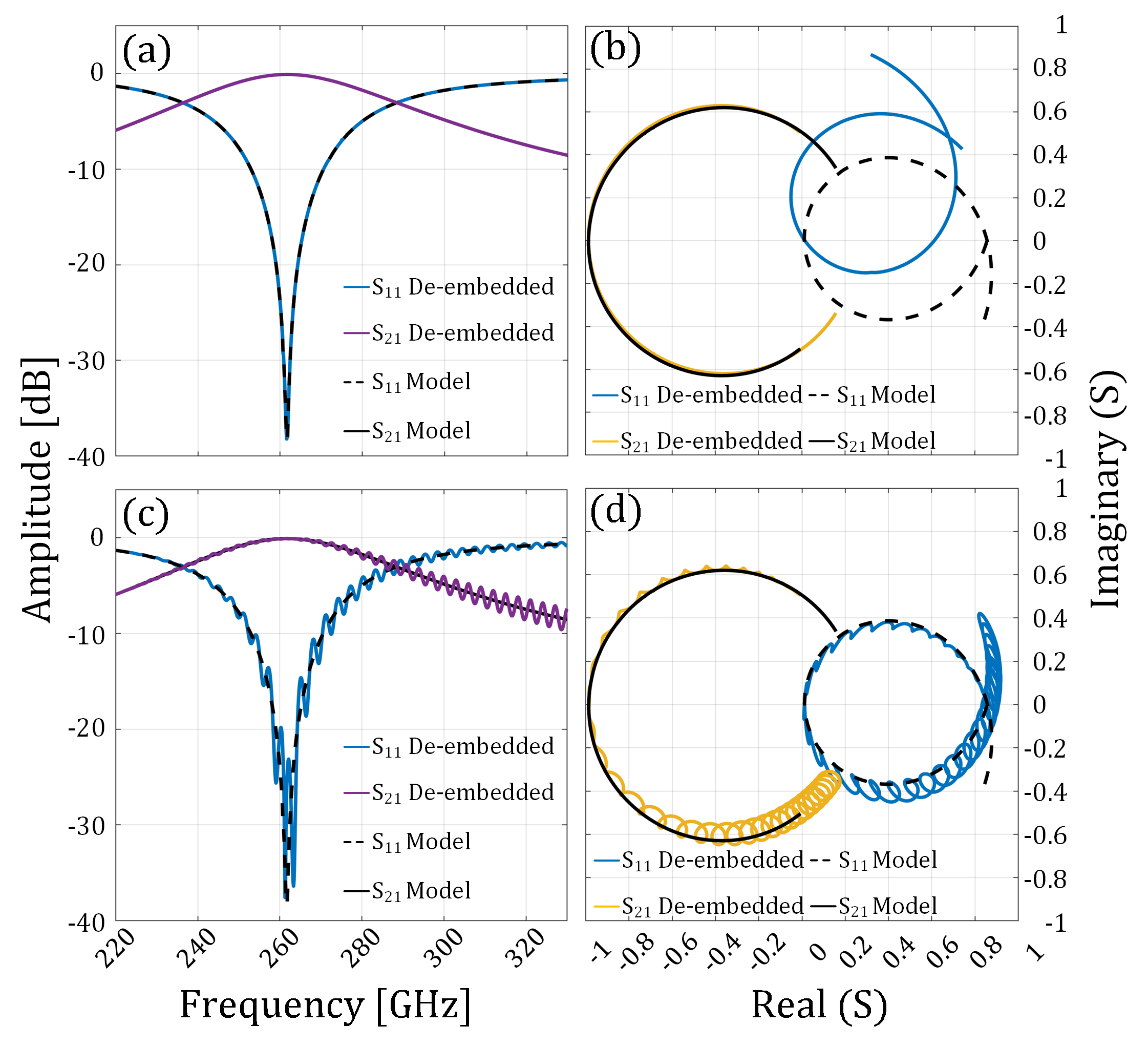}
    \caption{S-parameters in two most obvious error configurations: (a) displacement of the reference plane inside the cryostat by $5~\mu m$ and its impact on the (b) dependence of the real and imaginary parts of S-parameters on each other; (c) displacement of both optical windows towards the DUT by $5~\mu m$ and (d) corresponding imaginary vs. real part of S-parameters. Stainless steel filter numerically simulated S-parameters are used for the analysis.}
    \label{fig:explanation2}
\end{figure}

To investigate the impact of varying DUT movements, we performed numerical embedding of the stainless steel CST-simulated S-parameters into an initial error network, calculated the error network ABCD matrices, and then slightly disturbed it to see the consequences. In this calculation, we do not consider different error paths, e.g., cross-talk, leakage, etc., and consider only ABCD matrices for a quasioptical fixture and corresponding calibration de-embedding errors. 

Thus, a typical overall ABCD-matrix for an array of  materials with refractive indices $n_i=\sqrt{\epsilon_i}$ and thickness $t_i$ will be a product of separate matrices

\begin{equation}
\boldsymbol{ABCD}=\prod_i\boldsymbol{ABCD}_i=\prod_i 
 \begin{bmatrix}
\cos \beta_i t_i & jZ_0\sin \beta_i t_i \\
j Z_0^{-1}\sin \beta_i t_i & \cos \beta_i t_i 
\end{bmatrix} ,
\end{equation}

where $\beta_i=\frac{2\pi f n_i}{c}$ is the propagation constant in $i$-th medium, $c$ is the speed of light in vacuum, and $i$ corresponds to e.g., free space, then vacuum windows, interspace between them, and  DUT, in appropriate order.

In Fig.~\ref{fig:explanation2} (a), the device holder with a stainless steel filter has been shifted by 5 $\mu m$ inside the cryostat after the calibration. The corresponding ABCD matrices were calculated and translated into S-parameters. In this scenario, the internal distance between two mirrors remains the same $d_1 + d_2$, while the DUT surface is translated to the right (along the optical axis). While there is significant variation in the complex representation arising form displacement induced phase shifts, there are no visible effects on the the S-parameter magnitudes.

In Fig.~\ref{fig:explanation2} (b), the reference plane of the DUT remains the same, while the vacuum windows are shifted inside the cryostat by $5~\mu m$ each, i.e. $d_1+d_2$ decreases by $10~\mu\text{m}$, meanwhile $l_1=l_2$ increases by $5~\mu \text{m}$, leading to a redistribution of the standing wave patterns inside the cryostat. While this effect is not as observable at the small frequencies, we see that it grows as a function of frequency. This is quite understandable, since the wavelengths at higher frequencies get shorter, and then measurements are more sensitive to interference.

These perturbations can be handled with additional improvements dedicated to improving the system's stability. Thus, to prevent different error network perturbation effects, a more advanced calibration method would involve employing a stepper motor to replace samples inside the cryostat. In this setup, the stepper motor would control the sample holder, enabling calibration under both vacuum and cryogenic conditions. For instance, to measure the S-parameters of a DUT, the stepper motor could first replace the DUT with calibration standards for measurement, after which the VNA could compute the error matrix specific to the desired temperature and measurement. 

Additional improvement can be done by replacing the standard vacuum windows with ultrathin alternatives, such as Mylar-based windows. Kerr et al.\cite{kerr1992study} proposed a promising design using a Mylar window supported by a polystyrene layer. The polystyrene not only provides structural support and protection for the Mylar but also acts as an infrared radiation filter, a significant advantage when testing superconducting devices.

\section{\label{sec:conc}Conclusion}

In this paper, we have proposed a novel approach for cryogenic millimeter-wave measurements, enabling the testing of cryogenically cooled devices while de-embedding their performance from room-temperature quasi-optical components. We designed, simulated, and constructed a quasi-optical, cryogenically cooled, VNA-based two-port characterization system operating in the 220–330 GHz frequency range. The system's two-port calibration was performed using line, reflect, and match calibration standards. After characterizing the beam quality with a near-field scanner, we conducted measurements on three devices under test. The system demonstrated reliable performance for characterizing cryogenically cooled materials and devices in the 220–330 GHz range and holds promise for application in radio astronomy instrumentation testing.

\section*{Acknowledgements}

The authors would like to thank Dr. D. Cunnane, Dr. G. Chattopadhyay, Mr. J. Greenfield (Jet Propulsion Laboratory, Caltech) and Dr. H.  
Bohuslavskyi (VTT) for fruitful, stimulating discussions on cryogenic measurements and superconductors, Mr. J. M. Herrera Martin (Universidad Carlos III de Madrid) for assistance with measurements, and Dr. Sergei Kosulnikov (Aalto University at the time of the contribution, now Optenni) for numerical simulation suggestions. Easy-Cad Oy is warmly acknowledged for fabricating the stainless steel filters.

\bibliography{Ref}
\bibliographystyle{IEEEtran}

\end{document}